\documentclass{appolb}
\usepackage{epsfig}

\begin{document}
\title{Fluctuations and initial state granularity in heavy ion collisions
 and their effects on observables from hydrodynamics
\thanks{Presented at IV WPCF, Krakow 09/2008}
}
\author{R.P.G.Andrade, A.L.V.R.dos Reis,F.Grassi, Y.Hama,W.L.Qian
\address{Instituto de F\'{\i}sica-Universidade de S\~ao Paulo}
\and{T.Kodama}
\address{Instituto de F\'{\i}sica-Universidade Federal do Rio de Janeiro}
\and{J.-Y.Ollitrault}
\address{Institut de Physique Th\'eorique-Saclay}
}
\maketitle

\begin{abstract}
A comparison is made between results obtained using smooth initial conditions
and event-by-event initial conditions in the
hydrodynamical description of relativistic nuclear collisions.
Some new results on directed flow are also included.

\end{abstract}
  
\section{Objective}

Hydrodynamics has been rather successful at describing data obtained in 
relativistic nuclear collisions at RHIC.
Usually, smooth initial conditions are assumed (see e.g. fig.1 in \cite{hi}
and fig.3 in \cite{chi}. 
On the other side, microscopic codes such as NeXus predict initial conditions
event-by-event, which are quite irregular as shown in fig.1.

The question we address here is whether such structures (hot spots or more 
precisely hot tubes) can 
have a sizable effect on variables.

\begin{figure}[htbp]
\begin{center}
\epsfig{file=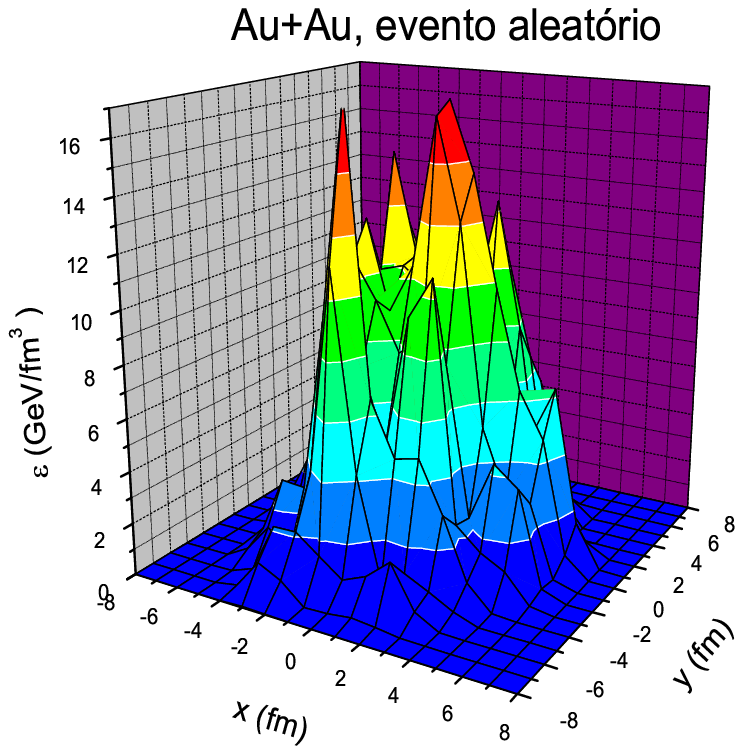,height=4.2cm}\hspace*{1.5cm}
\epsfig{file=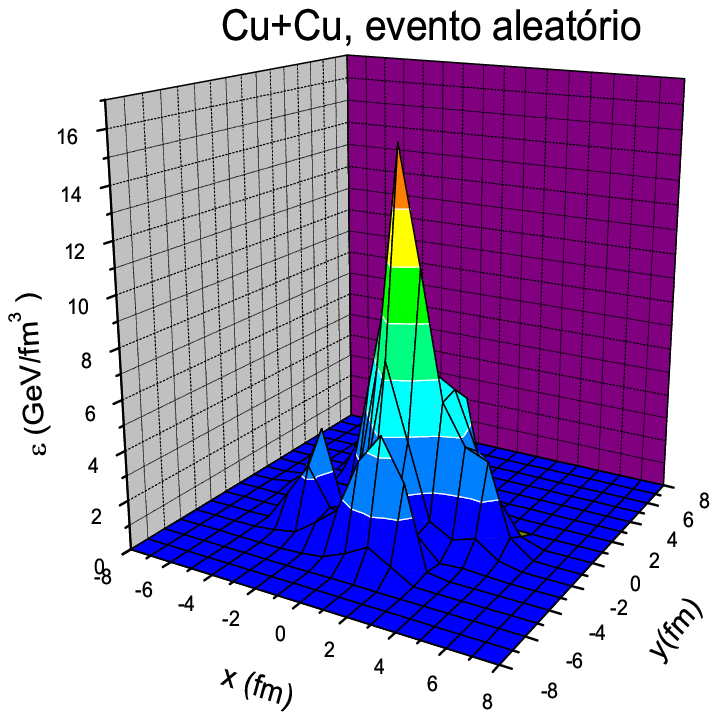,height=4.2cm}
\end{center}
\caption{$\eta=0$ slice for initial energy density of a RHIC collision in the 
6-15 \% centrality window.}
\end{figure}

To solve the hydro equations with very irregular initial conditions, we use the SPheRIO code. This code is based on the method of Smoothed Particle 
Hydrodynamics, originally developed in astrophysics and adapted to 
relativistic heavy ion collisions in \cite{SPH}.
The version of NeXSPheRIO used here has
initial conditions provided by NeXus \cite{IC}
and normalized 
    by an $\eta$-dependent factor to reproduce $dN_{ch}/d\eta$
     in each centrality 
    window \cite{norm}.
 The equation of state has a critical point \cite{crit}.
$T_{f.out}$ is fixed (mostly) by $dN_{ch}/p_tdp_t$ and depends on the 
centrality window (i.e. number of participants).
Centrality windows are defined 
using participant number and not impact parameter \cite{part}.
An ideal fluid is assumed,
a code with Smoothed Particle Hydrodynamics and dissipation is under 
development \cite{diss}.

\section{Comparison between fluctuating and 
average IC}

In the following, we present a summary of results  obtained using smooth initial conditions and running once the SPheRIO hydro code (standard approach)
or using a set of NeXuS
initial conditions, running for each initial conditions the  SPheRIO hydro code and computing 
averages over the set
for various observables (event-by-event hydrodynamics).

\subsection{$p_t$ distribution}
As can be seen in figure 2 (left),
the high $p_t$ part  is lifted. This is expected since hot tubes must expand more violently, producing more high $p_t$ particles \cite{PRL08,YH}.

\begin{figure}[htbp]
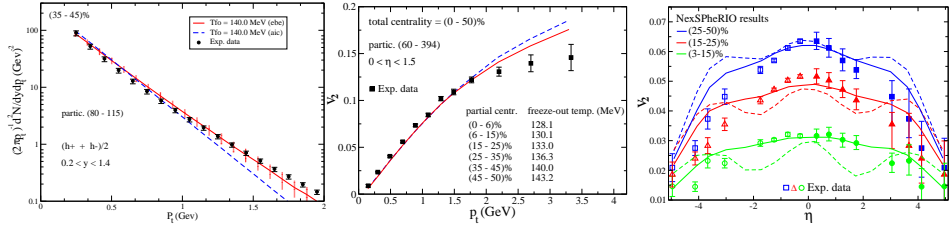

\begin{center}
\epsfig{file=dndpt1.eps,height=2.95cm}\epsfig{file=Rfig_4.eps,height=2.95cm}
\epsfig{file=Rfig_5.eps,height=2.9cm}
\end{center}
\caption{Left: charged particle $p_t$ distribution. Solid line: e-by-e initial 
conditions. Dashed: smooth initial conditions. Data: \cite{PHOBOS1}.
Center: $p_t$ dependence of $\langle v_2 \rangle$. Data: \cite{PHOBOS2}.
Right: $\eta$ dependence of $\langle v_2 \rangle$. Data: \cite{PHOBOS2}. }
\end{figure}

\subsection{elliptic flow}

$v_2(p_t)$ is flatter as seen in figure 2 (centre). This is also expected as the
isotropic expansion of hot tubes produces more high $p_t$ particles and lowers 
$v_2(p_t)$  \cite{PRL08,YH}.
In addition, $v_2(\eta)$ has no shoulder \cite{PRL06}
 as seen in figure 2 (right). 
The effects (isentropic expansion) of the hot tubes are more visible in regions of lower matter 
density present at larger $\eta$'s \cite{PRL08,YH}.


\subsection{Other comparisons}

In \cite{PRL04}, we argued that the hot tubes should manifest themselves giving smaller HBT radii. However, the situation might be more complicated.

Another observable where hot tubes might manifest themselves
is the ridge, a 
structure observed in the 2 particle correlations, plotted as function of
pseudorapitity difference $\Delta \eta$
and azimutal angle difference $\Delta \phi$
 between a high $p_t$ trigger hadron and its associated hadrons (see e.g. \cite{STAR}). The structure is
 $\Delta \eta$ independent. In NeXSPheRIO, the hot tubes can lead to such a ridge for the e-by-e initial conditions and not the smooth ones
\cite{ridge}.

Finally, the fluctuations in the e-b-e initial conditions also manifest 
themselves in fluctuations of $v_2$ (as well as $v_1$). The predicted
values for $v_2$ at 130 A GeV \cite{FLUC1}  and estimates at 200 A GeV 
\cite{norm} 
are in agreement with data \cite{FLUCPH,FLUCST}. 
Improvements to remove the non-flow effects have been reported by STAR and 
PHOBOS, see e.g. \cite{FLUCWPCF}.

\section{New results on directed flow}

In this section, we present some new {\it preliminary} results obtained with
NeXSPheRIO on directed flow.

\subsection{What is directed flow and what is expected}

If a nucleus-nucleus collision  is a number of independent nucleon-nucleon 
collisions,
 the momentum distribution is isotropic.
If instead, it leads to thermalized matter in the overlap region,
 the momentum distribution is stretched along the impact parameter direction,
$v_2$ is a measure of this stretching (so teaches about IC, thermalization, 
etc). There is also the possibility that the momentum distribution be 
shifted/deformed towards one of the sides in the x-y plane,
$v_1$ is a measure of this shift. 


At some energy, a ``wiggle'' in $v_1(\eta)$ is predicted.
In some microscopical models such as RQMD and UrQMD, this could be the case for 
nucleons at RHIC energy \cite{RQMD,UrQMD}.
In hydro models, this could be the case for the fluid, if a QGP phase occurs
\cite{v1Cs,v1Br,v1Ri,v1Iv}.
 
At SPS energy (40 A GeV and 158 A GeV), it was shown \cite{v1NA49} that
 pions and protons behave oppositely.
Pion directed flow as function of rapidity has no wiggle and crosses y=0 with a negative slope while 
nucleon directed flow has no wiggle and
crosses y=0 with a  positive slope
(except perhaps at the higher energy, in the more peripheral bin, where 
there is a hint of wiggle).

\subsection{RHIC results on directed flow}

At RHIC, directed flow for charged particles is rather similar to what was 
obtained  at SPS for pions: it crosses $\eta=0$ with a negative slope
\cite{v1PHOBOS,v1STAR1,v1STAR2}. 
This is understandable
since charged particles are mostly pions, the fluid directed flow must be dominated by pions. The turnover in $v_1(\eta)$ occurs for different values of $\eta$
in PHOBOS and in STAR (see below).

Results for identified particles are becoming available 
\cite{v1STAR3}. 

In addition, comparison of results for directed flow in 
Cu+Cu and Au+Au collisions show no system-size dependence \cite{v1STAR2}.

\subsection{NeXSPheRIO results on directed flow}

NeXSPheRIO results are in
qualitative agreement with PHOBOS for all $\eta$'s and
quantitative agreement for $\mid \eta \mid < 3 $ (figure 3 left).
They are in qualitative agreement with STAR for $\mid \eta \mid < 3$ but 
turnover occurs for smaller $\eta$ than for STAR (figure 3 right).

\begin{figure}[htbp]
\begin{center}
\epsfig{file=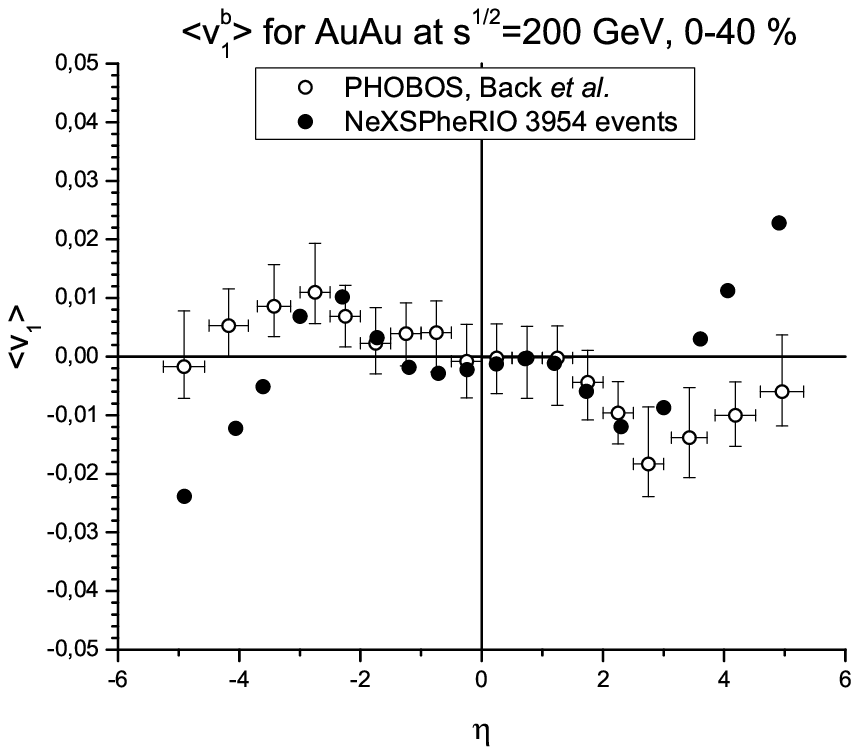,height=3.7cm}\hspace*{0.8cm}\epsfig{file=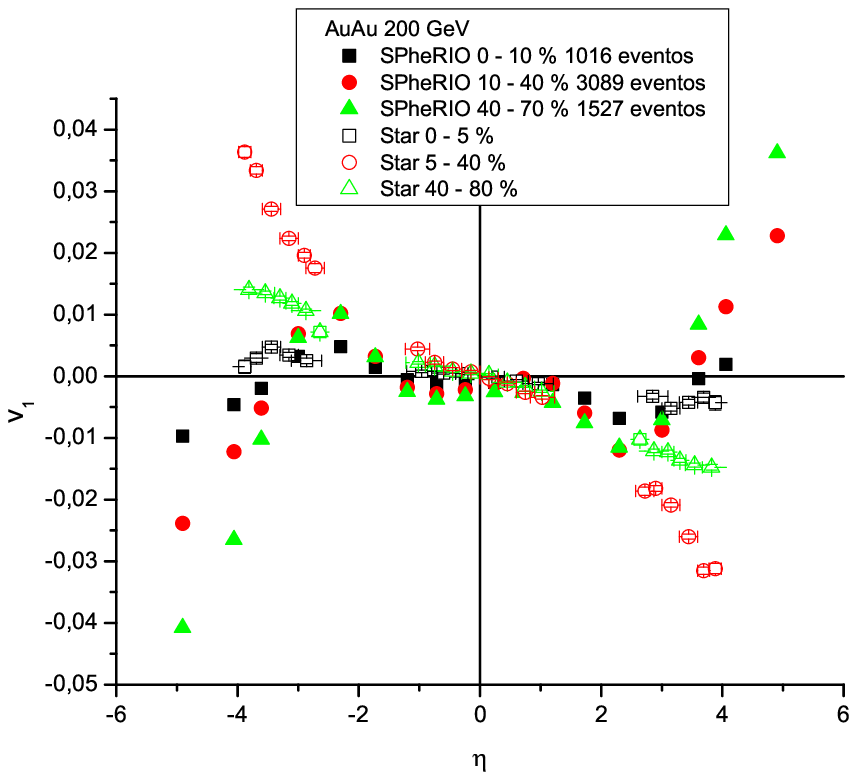,height=3.7cm}
\end{center}
\caption{Comparison of charged particle $\langle v_1 \rangle$ for NeXSPheRIO with
(left) PHOBOS \cite{v1PHOBOS} and (right) STAR \cite{v1STAR2}.}
\end{figure}


$v_1(\eta)$ from NeXSpheRIO for various 
centrality windows 
for Au+Au and Cu+Cu at 200 A GeV is shown in figure 4.
Little dependence on A is seen in the windows 6-15\% to 45-55\%.
Statistics must be improved.

\begin{figure}[htbp]
\begin{center}
\epsfig{file=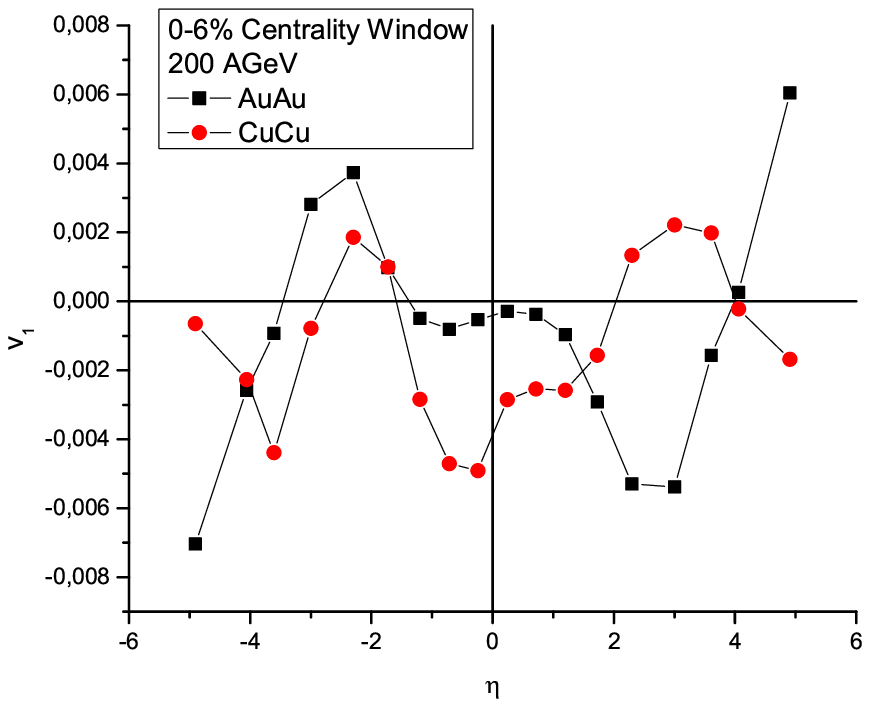,height=3.1cm}\epsfig{file=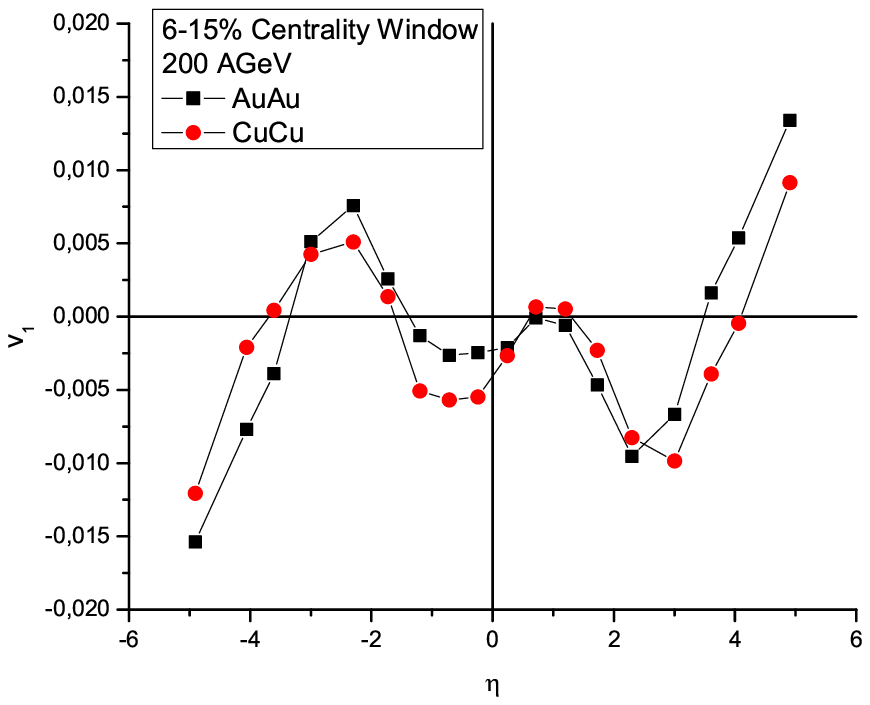,height=3.1cm}\epsfig{file=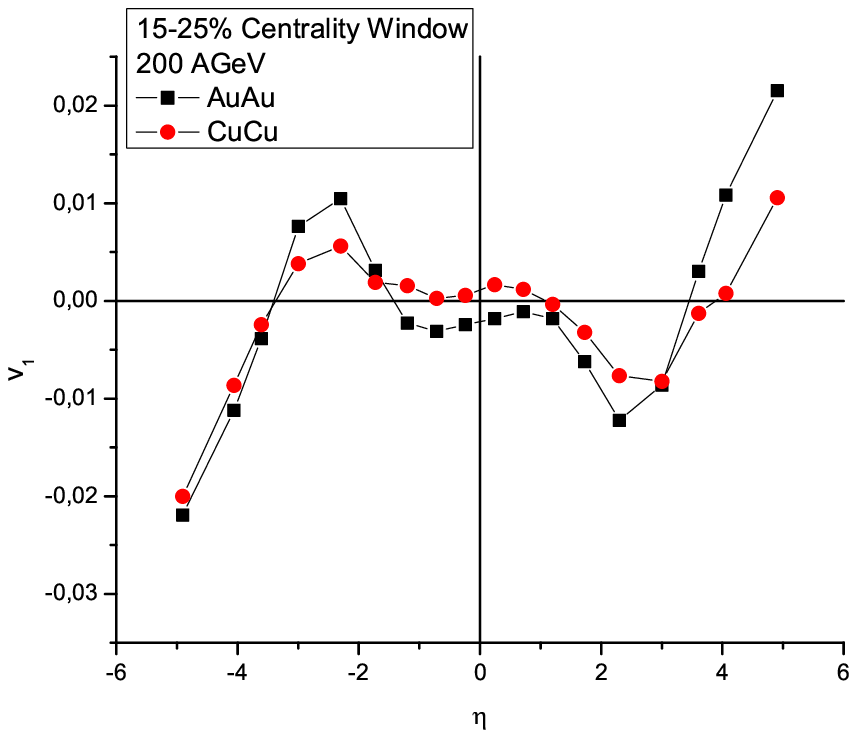,height=3.1cm}\\
\epsfig{file=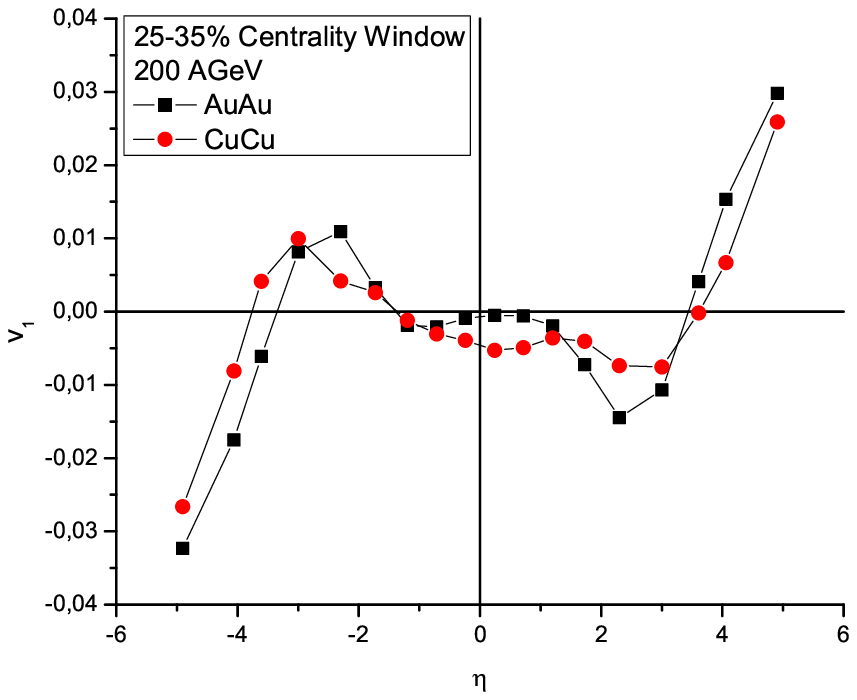,height=3.1cm}
\epsfig{file=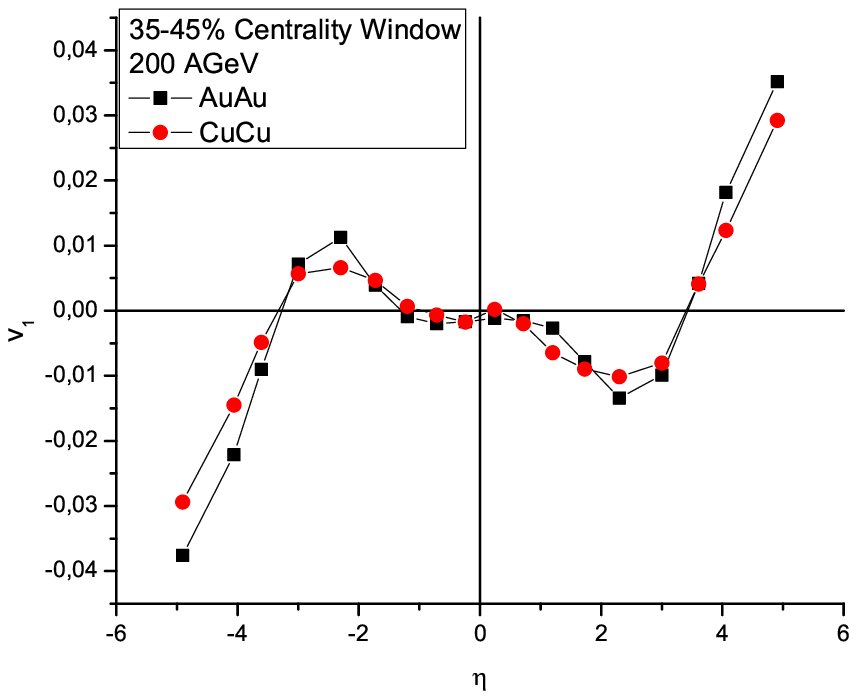,height=3.1cm}\epsfig{file=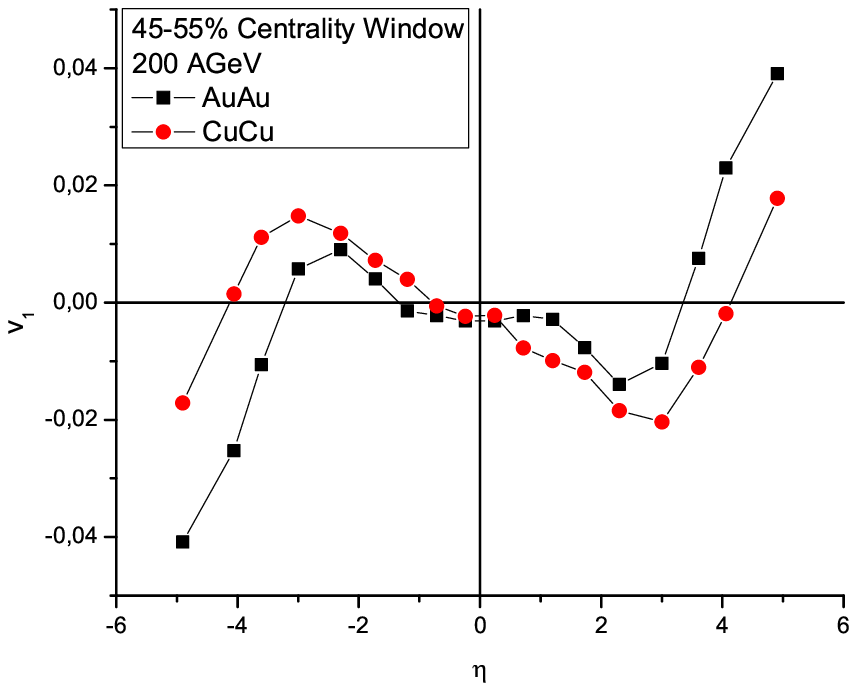,height=3.1cm}
\end{center}
\caption{Comparison of $\langle v_1 \rangle$ obtained in Cu+Cu and Au+Au, from NeXSPheRIO.}
\end{figure}

Figure 5 (left) illustrates particle dependence.
In NeXSPheRIO, protons have a big wiggle, pions have a plateau (left). 
A similar result was obtained using  UrQMD \cite{UrQMD}. 
In figure 5 (right), it is seen that $v_1(\eta)$ has a plateau for fluctuating 
initial conditions and a  somewhat stronger negative inclination for smooth 
initial conditions.

\begin{figure}[htbp]
\begin{center}
\epsfig{file=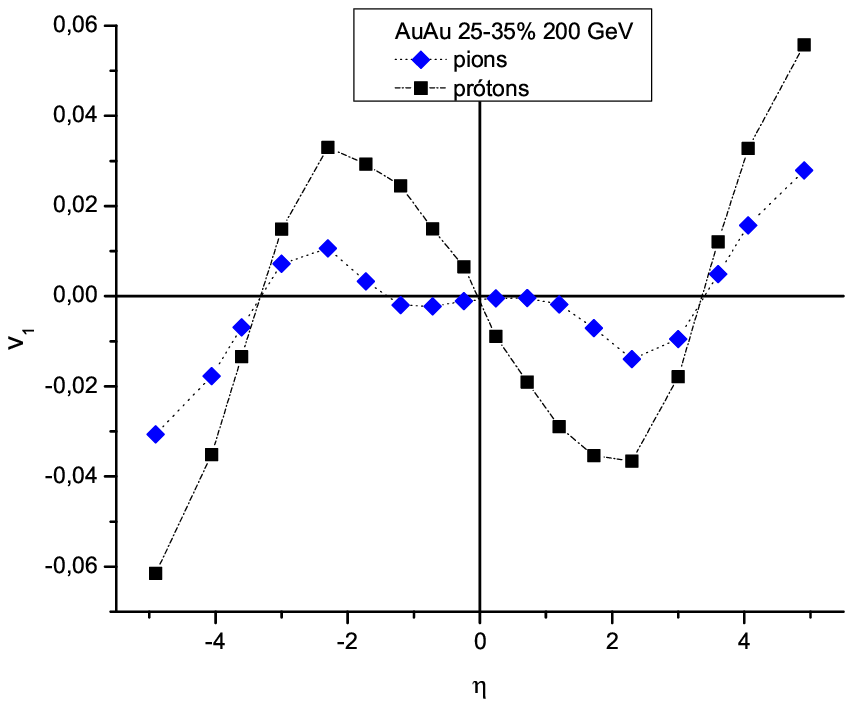,height=3.1cm}\hspace*{1.5cm}
\epsfig{file=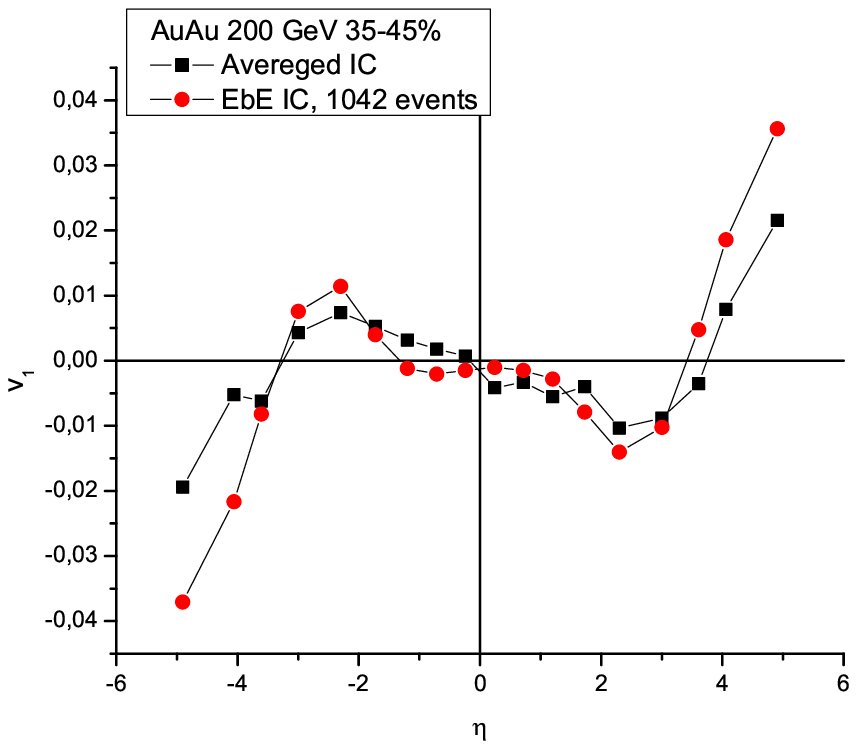,height=3.1cm}
\end{center}
\caption{Left: $\langle v1 \rangle$ for pions and protons. Right: $\langle v_1 \rangle$ for e-by-e and smooth initial conditions.}
\end{figure}



\section{Summary}

A short review of  possible effects of fluctuating initial conditions, rather than smooth ones, was presented. In addition to providing a reasonable description of various observables, as is possible with smooth initial conditions, some 
new effects were listed, most notably the ridge effect and the $v_2$ 
fluctuations, which do not appear when using the smooth initial conditions.

\end{document}